\documentclass[a4paper,11pt]{article}
\pdfoutput=1

\usepackage{array}
\usepackage{amsfonts}
\usepackage{amsmath}
\usepackage{amssymb}
\usepackage[small]{caption}
\usepackage{cite}
\usepackage[top=1.2in, bottom=1.2in, left=1.1in, right=1.1in]{geometry}
\usepackage{graphicx,subfigure}
\graphicspath{{./}{./figures/}}
\usepackage{pifont}
\usepackage{slashed}
\usepackage{color}
\usepackage{ulem}

\definecolor{mediumblue}{rgb}{0,0,0.8}
\usepackage{hyperref}
\hypersetup{
  linktocpage=true,
  colorlinks=true,
  citecolor=mediumblue,
  filecolor=mediumblue,
  linkcolor=mediumblue,
  urlcolor=mediumblue,
}
\newcommand{\mailref}[1]{\href{mailto:#1}{#1}}

\allowdisplaybreaks

\newcommand{\mc}{\mathcal}

\begin{document}

\begin{titlepage}
\def\thefootnote{\fnsymbol{footnote}}

\begin{flushright}
  \texttt{}\\
  \texttt{}
\end{flushright}

\begin{centering}
\vspace{0.5cm}
{\Large \bf \boldmath
  Novel Search for Light Dark Photon \\ 
in the Forward Experiments at the LHC}

\bigskip

\begin{center}
  {\normalsize 
    Yeong~Gyun~Kim$^{a,}$\footnote{\mailref{ygkim@gnue.ac.kr}},
    Kang~Young~Lee$^{b,}$\footnote{\mailref{kylee.phys@gnu.ac.kr, co-corresponding author}}, and
    Soo-hyeon~Nam$^{c,}$\footnote{\mailref{glvnsh@gmail.com, corresponding author}}}\\[0.5cm]
  \small
  $^a${\em Department of Science Education, Gwangju National
    University of Education,\\
    Gwangju 61204, Korea}\\[0.1cm]
  $^b${\em Department of Physics Education \& Research Institute of Natural Science,\\
  	Gyeongsang National University, Jinju 52828, Korea}\\[0.1cm]
  $^c${\em Department of Physics, Korea University, Seoul 02841, Korea}

\end{center}
\end{centering}

\medskip

\begin{abstract}
  \noindent
We propose a novel approach for discovering a light dark photon 
in the forward experiments at the LHC, 
including the SND@LHC and the FASER experiments. 
Assuming the dark photon is lighter than twice the electron mass and 
feebly interacts with ordinary matter,
it is long-lived enough to pass through 
100 m of rock in front of the forward experiments 
and also through the detector targets. 
However, some portion of them could be converted 
into an electron-positron pair inside the detector through their interaction with the detector target,
leaving an isolated electromagnetic shower
as a clear new physics signature of the dark photon. 
With copiously produced dark photons
from neutral pion decays in the forward region of the LHC,
we expect to observe sizable events inside the detector. 
Our estimation shows that more than 10 signal events 
of the dark photon could be observed 
in the range of kinetic mixing parameter, 
$6.2\times10^{-5} \lesssim \epsilon \lesssim 2\times10^{-1}$
and $3\times10^{-5} \lesssim \epsilon \lesssim 2\times10^{-1}$
for dark photon mass $m_{A^\prime} \lesssim$ 1 MeV
with integrated luminosities of 150 fb$^{-1}$ and 3 ab$^{-1}$, respectively.
\end{abstract}

\vspace{0.5cm}

\end{titlepage}

\renewcommand{\thefootnote}{\arabic{footnote}}
\setcounter{footnote}{0}

\setcounter{tocdepth}{2}
\noindent \rule{\textwidth}{0.3pt}\vspace{-0.4cm}\tableofcontents
\noindent \rule{\textwidth}{0.3pt}

\section{Introduction\label{sec:intro}}
%
The ATLAS and CMS detectors at the Large Hadron Collider (LHC) in CERN 
have been successfully operating and played pivotal roles 
in discovering Higgs boson in 2012\cite{higgs2012-atlas, higgs2012-cms}. 
While these detectors are well-equipped at large angles 
relative to the beamline and therefore
capable of detecting new physics with high transverse momentum $p_T$, 
they have holes in the far-forward direction. 
This limitation could potentially result in 
missed opportunities to detect (new) particles 
produced in the far-forward region.

To address this limitation, 
the FASER\cite{FASER:2018} and the SND@LHC\cite{SND:2021} experiments 
were designed to detect high-energy neutrinos 
and explore new particles that are produced in the far-forward direction 
at the LHC.
These experiments are located 480 m downstream 
of the ATLAS interaction point (IP) along the beam collision axis.
Charged particles from the IP are deflected by LHC magnets, 
while the remaining particles are further shielded 
by the 100 m of rock and concrete in front of the detectors. 
However, neutrinos and feebly interacting new particles 
would easily pass through this shielding and reach the detectors.

An example of new particles that can be probed in the forward experiments 
is a dark photon $A^\prime$, 
a hypothetical particle which interacts electromagnetically 
with standard model particles, 
but with a suppressed interaction strength $\alpha^\prime$ 
compared to ordinary fine structure constant $\alpha$.
If the dark photon mass $m_{A^\prime}$ is small enough, 
the main production channel of the dark photons at the LHC 
is through the decays of mesons, such as neutral pion $\pi^0$. 
Then, a significant number of dark photons 
would be produced in the far-forward direction, 
depending on the interaction strength.
If the mass of the dark photon is less than twice the electron mass,
the dark photon is long lived.
With the suppressed interaction strength $\alpha^\prime \ll \alpha$,
the majority of the produced dark photons would pass through 
the 100 m of rock in front of the forward experiments and reach the detector. 

When the light dark photon passes through detector target material 
(i.e. tungsten for the forward experiments), 
it can be converted into a pair of electron and positron, 
similarly to ordinary energetic photon, 
but with a significantly suppressed interaction rate.
Therefore, the signal signature of the dark photon in the forward experiments 
would manifest itself as an isolated electromagnetic (EM) shower 
with a shower shape essentially identical to that of an ordinary energetic photon.
If an ordinary photon enters the tungsten target, 
the EM shower will start near the detector entrance 
because the radiation length $X_0$ of tungsten is just 3.5 mm. 
On the other hand, 
if the EM shower of the dark photon occurs in the detector target, 
it would start at any location within the detector. 
This is because the `effective' radiation length of the detector material 
for the dark photon increases by a factor of 
$1/\epsilon^2\equiv\alpha/\alpha^\prime$. 
If $1/\epsilon^2 = 10^4$, for instance, 
the effective radiation length of tungsten
for the dark photon becomes 35 m, 
which is much longer than the actual detector target length
(i.e. $\sim$ 30 cm for the SND@LHC detector). 

In this paper, we will investigate the detection possibility 
of the light dark photon in the SND@LHC experiment at the LHC. 
In section 2, we introduce a dark photon model. 
We discuss dark photon production and detection 
in section 3 and 4, respectively. 
We conclude in section 5.

\section{A Dark Photon Model}
It might be possible that the standard model (SM) is accompanied 
by additional gauge structures, such as
a new $U(1)_X$ gauge group with a gauge field $X$. 
The gauge field $X$ can mix with the SM $U(1)_Y$ gauge field $Y$ through
a renormalizable kinetic mixing term $X_{\mu\nu}Y^{\mu\nu}$, 
where $X_{\mu\nu}$ and $Y_{\mu\nu}$ are the field strengths 
of $X$ and $Y$ gauge fields,
respectively.
If the SM particles are all uncharged under the $U(1)_X$, 
the relevant Lagrangian for the $X$ and $Y$ gauge fields, 
including a mass term for $X$,
can be written as
\begin{align} \label{eq:lagrangian}
	\mc{L} = -\frac{1}{4} X_{\mu\nu} X^{\mu\nu} 
	         -\frac{\sigma}{2} X_{\mu\nu} Y^{\mu\nu}
	-\frac{1}{4} Y_{\mu\nu} Y^{\mu\nu} + j^\mu_Y Y_\mu 
	+\frac{1}{2} m^2_X X_\mu X^\mu,
\end{align} 
where $j^\mu_Y$ denotes interaction current of 
gauge field $Y$.\footnote{
The Lagrangian in Eq.~(\ref{eq:lagrangian}) seems not manifestly gauge-invariant
due to the mass parameter of the $U(1)_X$ gauge boson, $m_X$.
Nonetheless, one can make it gauge-invariant and renormalizable by adopting the Stueckelberg action
which has an additional axionic scalar \cite{Stueckelberg:1938hvi}.  
Then, the topological mass parameter $m_X$ denotes the Stueckelberg coupling of the axionic scalar 
to the $U(1)_X$ gauge boson,
and our Lagrangian corresponds to the case that the Stueckelberg coupling to the $U(1)_Y$ gauge boson
is zero in the Stueckelberg extension of the SM with kinetic mixing \cite{Feldman:2007wj}.
}

In order to obtain the Lagrangian in the canonical form, 
we redefine the gauge fields as follows
\begin{align}
	\hat{Y}_\mu = Y_\mu + \sigma X_\mu, \,\,\,\,\,\,\, 
        \hat{X}_\mu=\sqrt{1-\sigma^2}\, X_\mu \,.
\end{align}
Then the Lagrangian is given by 
\begin{align}
	\mc{L} = -\frac{1}{4} \hat{X}_{\mu\nu} \hat{X}^{\mu\nu}
	-\frac{1}{4} \hat{Y}_{\mu\nu} \hat{Y}^{\mu\nu} 
	+ j^\mu_Y \big(\hat{Y}_\mu 
	   - \frac{\sigma}{\sqrt{1-\sigma^2}} \hat{X}_\mu\big)
	+\frac{1}{2} \frac{m_X^2}{(1-\sigma^2)}\hat{X}_\mu\hat{X}^\mu \,.
\end{align}
Here we will focus only on the EM interaction in the SM. 
If we identify the massless gauge boson $\hat{Y}$ 
as the SM photon field $A$ and
the massive gauge boson $\hat{X}$ as the dark photon field $A^\prime$,
the resulting Lagrangian can be written as \cite{Fabbrichesi:2020wbt}
\begin{eqnarray}
	\mc{L}=-\frac{1}{4} F_{\mu\nu} F^{\mu\nu}
	-\frac{1}{4} F^\prime_{\mu\nu} F^{\prime\mu\nu}
	+\frac{1}{2} m^2_{A^\prime} A^{\prime 2}
	+j^\mu_{em} (A_\mu + \epsilon A^\prime_\mu),
\end{eqnarray}
where $F_{\mu\nu}$ and $F^\prime_{\mu\nu}$ 
are the field strengths of the photon $A$ and 
dark photon $A^\prime$, respectively,
$m_{A^\prime}$ is the mass of the dark photon, and 
$j^\mu_{em}$ is the EM current of the SM particles. 
The dark photon couples to the EM current
with a coupling suppressed by $\epsilon$ compared to the ordinary photon coupling. 
Therefore, we have two new free parameters $m_{A^\prime}$ and $\epsilon$
in this dark photon model, in addition to the SM parameters. 
The Lagrangian in Eq.(4) is the starting point of our analysis.

\section{Dark Photon Production}
The LHC at 13 TeV center of mass energy has 
a total inelastic proton-proton scattering cross section of around 75 mb, 
as reported by ATLAS and CMS Collaborations \cite{ATLAS:2016, CMS:2016}. 
It implies the production of large number of particles 
that are highly concentrated 
in the far-forward direction. 
For example, each hemisphere is expected to have approximately 
$n_{\pi^0} \approx 2.1 \times 10^{17}$ neutral pions 
for the 150 ${\rm fb}^{-1}$ of integrated luminosity. 
Roughly 11$\%$ of all neutral pions in one hemisphere 
are produced within the pseudo-rapidity range 
($7.2 < \eta <8.4$) for the SND@LHC experiment.
\begin{figure}
	\centering
	\subfigure[$\pi^0$ production by EPOS-LHC]{\label{fig:pi0_epos} %
	\includegraphics[height=5.7cm]{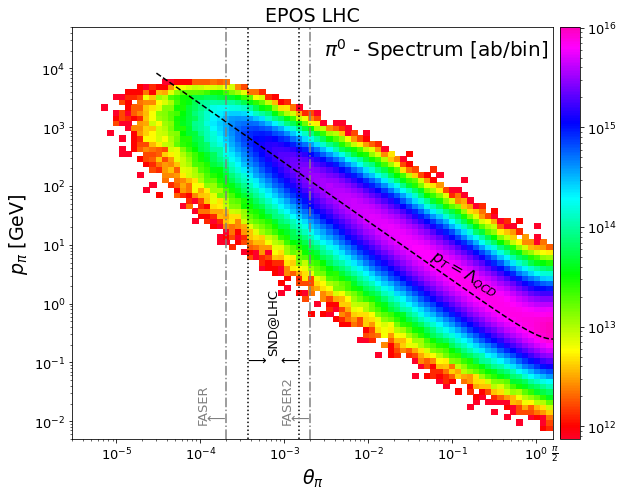}} \
	\subfigure[$\gamma$ production by EPOS-LHC]{\label{fig:gamma_epos} %
	\includegraphics[height=5.7cm]{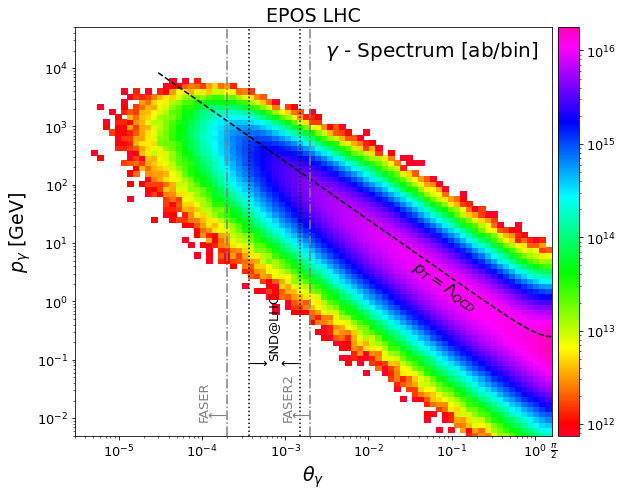} } \\
	\subfigure[$\pi^0$ production by PYTHIA8]{\label{fig:pi0_pythia} %
	\includegraphics[height=5.7cm]{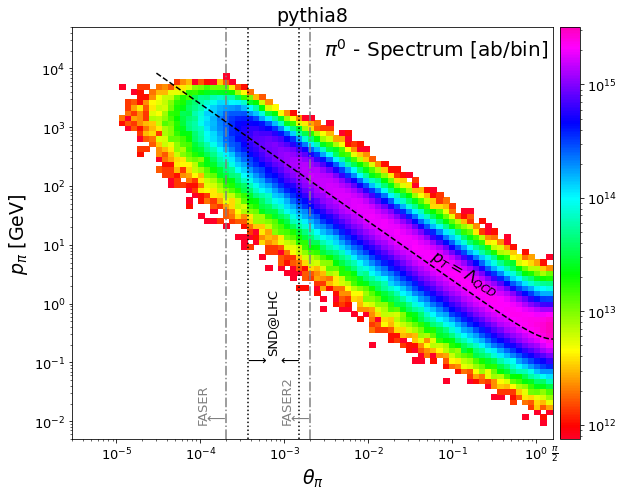}} \
	\subfigure[$\gamma$ production by PYTHIA8]{\label{fig:gamma_pythia} %
	\includegraphics[height=5.7cm]{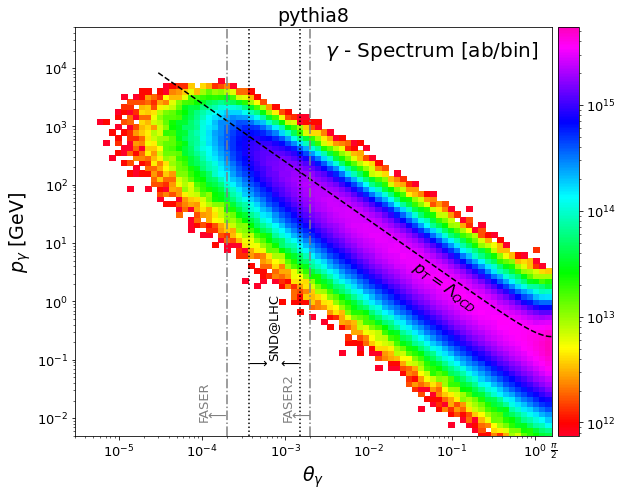} }
	\caption{ (a) Neutral pion and (b) photon production rates in each hemisphere on the plane of 
		the production angle $\theta$ and the momentum $p$ of the particles.
		The spectra were obtained with the Monte Carlo generator EPOS-LHC implemented
		in the CRMC simulation package. Also shown are the results with PYTHIA8 for (c) neutral pion
		(d) photon production rates.
	}
\end{figure}
Fig.~1(a) shows the neutral pion production rate in one hemisphere 
in the $(\theta_\pi, p_\pi)$ plane,
where $\theta_\pi$ and $p_\pi$ are the meson's production angle with respect to the beam axis and momentum,
respectively. The neutral pion spectrum was obtained with the Monte Carlo generator EPOS-LHC\cite{epos-lhc} 
implemented in the CRMC simulation package\cite{crmc}.
One can see Ref.\cite{faser1708, forsee} for a similar previous analysis.

If the dark photon mass is smaller than neutral pion mass, the dominant production channel of the dark photons is neutral pion decay, 
$\pi^0 \rightarrow A^\prime \gamma$.
The branching ratio of the decay mode is given by\cite{batell}
\begin{eqnarray} \label{branching}
	B(\pi^0 \rightarrow A^\prime\gamma) = 
	2\epsilon^2\left(1-\frac{m_{A^\prime}^2}{m_{\pi_0}^2}\right)^3 
	B(\pi^0 \rightarrow \gamma\gamma),
\end{eqnarray}
with $B(\pi^0 \rightarrow \gamma\gamma) \simeq 0.99$. 
%
The neutral pion is a pseudoscalar, and so (dark) photons are produced isotropically
in the meson's rest frames.
Fig.~1(b) shows photon production rate in one hemisphere in the ($\theta_\gamma, p_\gamma$) plane.
In the figure, in addition to the pion decay, other production channels of photons 
such as $\eta$ meson decay, proton bremsstrahlung and Drell-Yan process etc. are also included.  
The number of photons in each hemisphere for 150 fb$^{-1}$ integrated luminosity is 
$n_\gamma \approx 4.5 \times 10^{17}$. About 10$\%$ of photons in one hemisphere are produced
in the pseudo-rapidity range of the SND@LHC detector. Considering also the geometrical acceptance
of the SND@LHC detector, it implies that the number of photons in the direction of
the SND@LHC detector volume is $n_\gamma^{\rm SND} \approx 10^{16}$.
If the dark photon mass is small enough (say, $m_{A^\prime} \lesssim 1$ MeV),
the spectrum of dark photons is essentially the same as the one of photons, and
the number of dark photons in the direction of the SND@LHC detector
scales as 
\begin{eqnarray}
   n_{A^\prime}^{\rm SND} = \epsilon^2 \times n_\gamma^{\rm SND} 
   \approx \epsilon^2 \times 10^{16} 
\end{eqnarray}
assuming the integrated luminosity of 150 fb$^{-1}$.
Note that around 99$\%\, (65 \%)$ of the dark photons produced along the SND@LHC direction have
energies larger than 10 (100) GeV.
We also show the simulation results with PYTHIA8 in Fig.~1(c) and 1(d) 
for the neutral pion and photon, respectively. 
In this case, the number of photons in the direction of
SND@LHC detector is similar to EPOS-LHC prediction, but slightly less, i.e. 
$n_\gamma^{\rm SND} \approx 0.6 \times 10^{16}$. 
We will proceed with the EPOS-LHC prediction for our analysis.

\section{Dark Photon Detection}
%
When photons pass through matter, the total number of photons 
is reduced by the number of interacted photons.
The attenuation of photon beam is exponential 
with respect to travel distance inside absorber, 
\begin{eqnarray}
	I(x) = I_0 \, {\rm exp}(-\mu x),
\end{eqnarray}
where $I_0$ is incident beam intensity, $x$ travel distance.
The absorption coefficient $\mu$ is given by $\mu = N \sigma$,
where $N$ is the density of atoms of the traversed material
and $\sigma$ is the total cross-section per atom.
Electron-positron pair production is the dominant mode of photon interaction at high photon energies.
In this case, the absorption coefficient $\mu$ is given by 
$\mu = N\tau_{pair}$ with $\tau_{pair}$, the pair production cross section.
For tungsten, the pair production cross section $\tau_{pair}$ 
is about 35 barns/atom and
the absorption coefficient is given by
\begin{eqnarray}
\mu = N\tau_{pair} \simeq 2.2/{\rm cm} \,\,\,\,({\rm tungsten}).
\end{eqnarray}
The absorption coefficient $\mu$ is related to the radiation length $X_0$ 
of the absorber as follows: 
$X_0 \simeq 7/9\mu$, which gives $X_0 \simeq 0.35\,{\rm cm}$ 
for tungsten material.

The light dark photon with $m_A \lesssim {\rm 1\, MeV}$ is predominantly produced from the neutral pion decay,
$\pi^0 \rightarrow A^\prime\gamma$. 
Such dark photon is transversely polarized because the co-produced photon is massless. 
One can see also from Eq. (\ref{branching}) that
there is no longitudinal polarization contribution to $B(\pi^0 \rightarrow A^\prime\gamma)$ 
which is rescaled by the factor of $\epsilon^2$ compared to $B(\pi^0\rightarrow \gamma\gamma)$ 
apart from a small kinematic mass factor. 
Furthermore, in its scattering with the SND@LHC detector material, the dark photon is very energetic 
so that its (kinematic) mass effects on the process is negligible.
Therefore, we can approximate that the cross section of electron-positron pair production by the light dark photon
is rescaled by a factor of $\epsilon^2$ compared to the photon case. 
The absorption coefficient $\mu$ 
for the dark photon beam is reduced by $\epsilon^2$ accordingly.
The photons and dark photons produced in forward direction 
at the ATLAS IP encounter the 100 m of rock 
in front of the forward experiment detectors. 
Therefore, we need to know the absorption coefficient for the rock. 
The average rock density around CERN is measured to be 
about 2.5 $\rm g/cm^3$\cite{Fern18}.
The rock can be modeled as a mixture of 41$\%$ $\rm CaCO_3$
and 59$\%$ $\rm SiO_2$ \cite{FASER:2021mtu}.
The radiation lengths are $X_0 \simeq 12.3 \,\rm cm$ for $\rm SiO_2$ and 
$X_0 \simeq 8.6 \,\rm cm$ for $\rm CaCO_3$.
From general formula for the radiation length of composite materials, 
we can obtain the radiation length $X_0 \simeq 10.4 \,{\rm cm}$ for the rock.
In turn, it gives the absorption coefficient $\mu$ for the rock,
\begin{eqnarray}
	\mu \simeq \frac{7}{9}\times\frac{1}{10.4\,\rm cm} \simeq 0.075/{\rm cm}\,\,\,\,({\rm rock}).
\end{eqnarray}
For energetic photons, the attenuation factor ${\rm exp}(-\mu x)$ for the 100 m of the rock is estimated to be
\begin{eqnarray}
{\rm exp}(-0.075/{\rm cm} \times 10^4 \,{\rm cm}) \simeq 0. 
\end{eqnarray}
Therefore, all photons from the ATLAS IP are absorbed by the rock 
and cannot reach the forward experiment detectors.
However, in the case of dark photon, 
the absorption coefficient $\mu$ is reduced by $\epsilon^2$.
For instance, if we assume $\epsilon^2 = 10^{-6}$, 
the attenuation factor is given by
\begin{eqnarray}
	{\rm exp}(-\mu x) \simeq {\rm exp}(-10^{-6} \times 0.075 /{\rm cm} \times 10^4\, {\rm cm}) 
	\simeq 0.9925
\end{eqnarray}
which implies that $99.25\%$ of dark photons just pass through 100 m of rock
and reach the tungsten target of the forward experiment detectors. 
For the SND@LHC detector, total thickness of the tungsten target is 30 cm.
The interaction rates of dark photons inside the SND@LHC tungsten target is then given by
\begin{eqnarray}
	1-\rm exp(-\mu x) = 1-exp(-10^{-6} \times 2.2/cm \times 30\, cm) 
	\simeq 6.6\times 10^{-5}.
\end{eqnarray}
Therefore, the fraction of $6.6 \times 10^{-5}$ among the dark photons which enter the detector
will be detected by the electron-positron pair production inside the detector.
For $\epsilon^2 = 10^{-6}$, the number of dark photons which are produced at the ATLAS IP
in the direction of the SND@LHC detector is $n_{A^\prime}^{\rm SND} \simeq 10^{10}$
for 150 fb$^{-1}$, as can be seen from the Eq.(6). Then the number of dark photon events
inside the detector for 150 fb$^{-1}$ is estimated to be
\begin{eqnarray}
	n_{\rm event} \simeq 10^{10} \times 0.9925 \times 6.6 \times 10^{-5} \simeq 6.6 \times 10^5.
\end{eqnarray}
Thus, huge number dark photon events can be obtained with 150 fb$^{-1}$ integrated luminosity, for $\epsilon^2 = 10^{-6}$. For general $\epsilon$ values, assuming 150  fb$^{-1}$, the number of dark photon events in the SND@LHC detector is given by
\begin{eqnarray}
	n_{\rm event} = \epsilon^2 \times 10^{16} \times 
	{\rm exp}(-\epsilon^2 \times 0.075 \times 10^4) \times
	\big(1 - \rm exp(-\epsilon^2 \times 2.2 \times 30)\big)
\end{eqnarray}

Fig. 2 shows the number of dark photon events as a function of $\epsilon$.
If $\epsilon$ is too large, 
all dark photons will be absorbed by the rock before reaching 
the detector target. On the other hand, if $\epsilon$ is too small, 
all dark photons will just pass through the rock 
and the detector without leaving any detectable signals in the detector. 
Therefore, the forward experiments have sensitivity only for 
a certain range of $\epsilon$. 
If we require the number of events be larger than 10, 
the sensitive range of $\epsilon$ will be 
$6.2\times 10^{-5} \lesssim \epsilon \lesssim 2\times10^{-1}$ for an integrated luminosity of 150 fb$^{-1}$ obtained during the LHC Run 3 
and $3\times 10^{-5} \lesssim \epsilon \lesssim 2\times10^{-1}$ for 3 ab$^{-1}$ within HL-LHC precision.
\begin{figure}
	\centering
	\includegraphics[height=7.5cm]{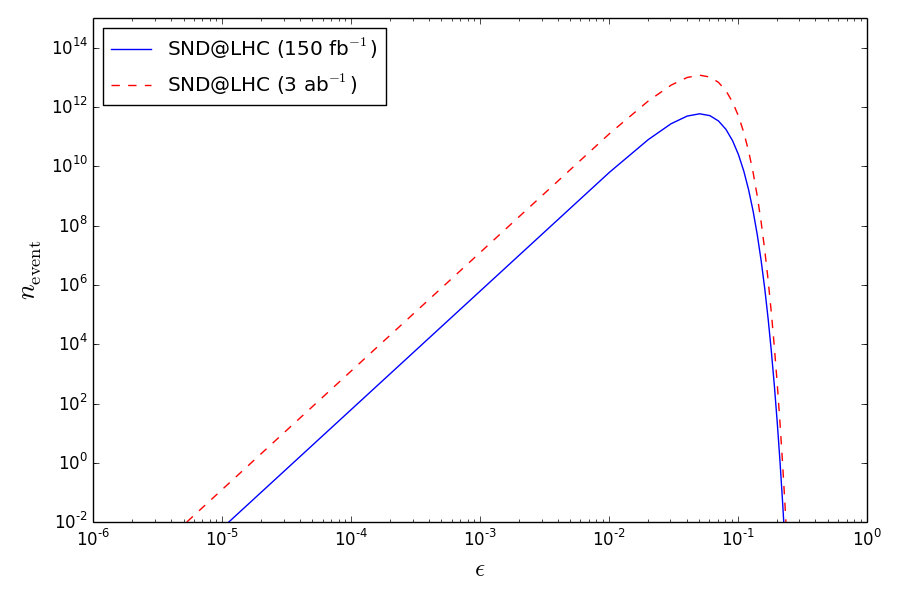}
	\caption{
	 The expected number of dark photon events as a function of $\epsilon$ 
	 with integrated luminosities 150 $\rm fb^{-1}$ (blue solid line) and 3 $\rm ab^{-1}$ (red dashed line).
	}
\end{figure}

Fig. 3 shows the sensitivity limit in the parameter plane ($m_{A^\prime}, \epsilon$) of the dark photon model from the SND@LHC experiment 
with integrated luminosities of 150 $\rm fb^{-1}$ and 3 $\rm ab^{-1}$ 
in comparison with other results such as atomic spectra\cite{atomicspectra}, TEXONO\cite{texono,park17} and 
EDELWEISS-III\cite{edelweiss}. 
Also shown are the bounds from electron anomalous magnetic moment $(g-2)_e$\cite{Pospelov}, 
Supernova 1987A using the ``robust" result\cite{SN1987A}.
There are also possible bounds in this mass region 
from the degrees of freedom ($\Delta \textrm{N}_\nu^\textrm{eff}$) counting in the early universe \cite{Delta_Neff}
and from neutrino experiment such as  LSND\cite{Pospelov18,LSND}.
However, we do not include those bounds 
because the counting of additional degrees of freedom would depend on details of the dark sector in the new physics model
and it is difficult to assess the exact signal process for Liquid scintillator neutrino detector as discussed in Ref.\cite{BEH}. 
The experimental sensitivity of the SND@LHC with 150 fb$^{-1}$ is stronger than that of the atomic spectra experiments
and a bit weaker than that of the TEXONO experiment.
Nonetheless, the experimental sensitivity of HL-LHC precision with 3 $\rm ab^{-1}$ is expected to be stronger than 
that of the TEXONO experiment.

\begin{figure}
	\centering
	\includegraphics[height=7.5cm]{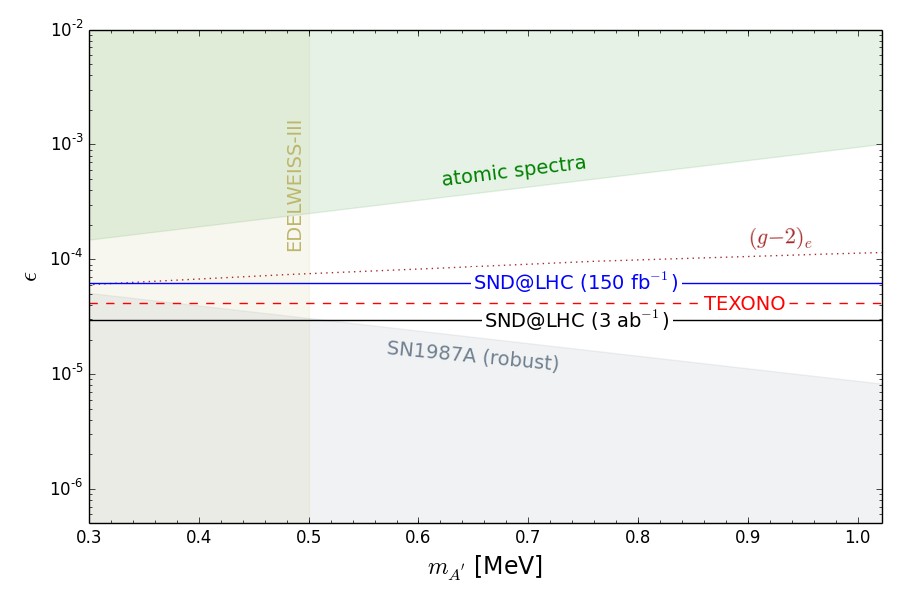}
	\caption{
		Sensitivity limit in the parameter plane ($m_{A^\prime}, \epsilon$) of the dark photon model from the SND@LHC experiment 
		with integrated luminosities 150 $\rm fb^{-1}$ (blue solid line) and 3 $\rm ab^{-1}$ (black solid line) 
		in comparison with other results such as atomic spectra (green shaded), TEXONO (dashed line) and EDELWEISS-III (khaki shaded).
		 Also shown are the bounds from electron anomalous magnetic moment $(g-2)_e$ (dotted line), Supernova 1987A (grey shaded).
	}
\end{figure}

At the detector level, the signature of dark photon signal is 
an isolated EM shower originated from electron-positron pair production 
by dark photon interaction with target material.
If it were originated from ordinary photon, 
the isolated EM shower would start at the entrance
of the detector because of the small radiation length $X_0$ of the detector target. 
But the starting points of the isolated EM showers 
originated from dark photons would appear to spread 
almost equally through detector target, 
due to the suppressed interaction strength
which gives suppressed absorption coefficient $\mu$ for dark photon.

A possible dominant background for the dark photon signal would be 
neutrino-electron scattering $\nu_x e^- \rightarrow \nu_x e^-$, which produces 
an isolated EM shower. 
However, it is estimated that the overall background yield is less than one 
for 150 fb$^{-1}$ integrated luminosity\cite{SND:2021}.
Furthermore, EM showers originating from (dark) photons are distinguished from
those initiated by electrons with a high efficiency close to 99$\%$.
This is mainly because the micrometric accuracy of the nuclear emulsion in the detector 
is capable of observing the displaced vertex associated with the photon conversion\cite{SND:2021}.
Another possible dominant background is neutral particles originating from primary muons interacting in
rock and concrete in front of the detector, which can potentially mimic the dark photon signal since
they can create a shower in the detector while leaving no incoming trace. 
However, it is shown that the rate of neutral hadron events with energies above 100 GeV
is heavily suppressed by using the veto system in the detector to tag the accompanying charged particles
(most often the scattered muon)\cite{SND:2023}. Therefore, we can easily remove the background 
by a selection cut on the energy of the shower.
In this regards, we assume that the backgrounds for dark photon signal is negligible.

Another possible way to detect dark photon at the forward experiments is through its decay to three photons,
 i.e. $A^\prime \rightarrow \gamma\gamma\gamma$. In the Euler-Heisenberg limit\cite{EH},
the dark photon decay width is given by
\begin{eqnarray}
	\Gamma_{\rm EH} = \frac{17\epsilon^2 \alpha^4}{11664000 \pi^3} \frac{m_{A^\prime}^9}{m_e^8} 
	\simeq 1\, {\rm s}^{-1} \bigg(\frac{\epsilon}{0.003}\bigg)^2 \bigg(\frac{m_{A^\prime}}{m_e}\bigg)^9
\end{eqnarray}
However, the exact width at one-loop order $(\Gamma_{\gamma\gamma\gamma})$ \cite{BEH}
shows substantial deviations from the result in the Euler-Heisenberg limit.
For instance, $\Gamma_{\gamma\gamma\gamma}/\Gamma_{\rm EH} \sim 10^2$ for $m_{A^\prime} = 2 m_e$.
The number of detected dark photons through the three photon decay is given by
\begin{eqnarray}
	n_{\rm event} = n_{A^\prime}^{\rm SND} 
	\bigg[{\rm exp}\bigg(-\frac{l}{\gamma\beta c\tau}\bigg) 
	- {\rm exp}\bigg(-\frac{l+\Delta l}{\gamma\beta c\tau}
	\bigg)\bigg]
\end{eqnarray}
where $c\tau$ is the proper decay length of dark photon, $l = 480$ m, and $\Delta l \simeq 0.3$ m.
For $m_{A^\prime} = 2 m_e$, the proper decay length is $c\tau \sim 0.05/\epsilon^2$ m.
If we further assume that $E_{A^\prime} = 100$ GeV and $\epsilon = 10^{-3}$, then $\gamma\beta c\tau \sim 5\times 10^9$ m
and $n_{\rm event} \simeq 0.6$. Therefore the sensitivity limit of the three photon channel is about 
$\epsilon \sim 10^{-3}$ at best.

\section{Conclusion}
In this work, we investigated detection possibility of light dark photon 
in the forward experiments at the LHC, 
especially the SND@LHC.
The dark photon interacts with the SM particles
through usual EM interaction but with suppressed strength by $\epsilon$ 
and can have an arbitrary mass $m_{A^\prime}$.
If dark photon mass is smaller than the neutral pion mass, 
the dominant production channel of dark photon is a neutral pion decay 
$\pi^0 \rightarrow \gamma A^\prime$.
A copious number of dark photon could be produced from the ATLAS IP,
while the exact number of dark photons depends mainly 
on the new parameter $\epsilon$. 
About $2\%$ of dark photons in each hemisphere 
would be produced within the geometrical acceptance range of the SND@LHC experiments. 
If the dark photon mass is smaller than twice of the electron mass, 
the dark photons are long-lived.
The majority of such dark photons could easily pass through 
the 100 m of rock in front of the forward experiments at the LHC, 
and reach the detector targets of the forward experiments. 
Although the most of dark photons entering the detector 
would just pass through the detector,
some portion of dark photons could be converted 
into electron-positron pair leaving an isolated EM shower 
as a new physics signature of the dark photon.
Our estimation shows that
more than 10 dark photon events would be produced 
in the range of $6.2\times 10^{-5} \lesssim \epsilon \lesssim 2\times 10^{-1}$
for dark photon mass $m_{A^\prime} \lesssim$ 1 MeV
with  integrated luminosity of 150 fb$^{-1}$,
and thus such a light dark photon can possibly be detected in the forward experiments at the LHC.

\section*{Acknowledgments}
This work is supported by Basic Science Research Program 
through the National Research Foundation of Korea (NRF) 
funded by the Ministry of Education under
the Grant No. RS-2023-00248860 (S.-h.N.)
and also funded by the Ministry of Science and ICT 
under the Grants No. NRF-2020R1A2C3009918 (S.-h.N.)
and No. NRF-2021R1A2C2011003 (K.Y.L.).

\end{document}